\documentstyle[12pt]{report}

\if@twoside
\else
\fi

\begin{document}
\newcommand{\be}{\begin{equation}}
\newcommand{\ee}{\end{equation}}

\vspace*{5mm}

\begin{center}  {\Large{Locally Disordered Lattices in Strong ac
Electric Fields}}

\vspace{20mm}

                {\large{Daniel W.~Hone}}

\vspace{5mm}

        Department of Physics and Center for Quantized Electronic
Structures \\
        University of California, Santa Barbara, CA 93106

\vspace{10mm}

                {\large{Martin Holthaus}}

\vspace{5mm}

        Department of Physics, Center for Nonlinear Sciences, \\
        and Center for Free--Electron Laser Studies \\
        University of California, Santa Barbara, CA 93106

\end{center}

\vspace{10mm}

\begin{description}

\item[Abstract.] ---
We develop a formalism to study the role of local defects
in tight binding systems in the presence of a strong external ac
electric field.

It is found that the appearance and disappearance of localized
states, as

well as their localization lengths, can be controlled by the
amplitude or
frequency of the driving field. The theory is illustrated by
numerical model
calculations for imperfect semiconductor superlattices in
far--infrared laser fields.
\end{description}

\vspace{5mm}

PACS numbers: 78.66.-w, 71.50.+t, 73.20.Dx

\vfill
\break

\setcounter{chapter}{1}
\setcounter{equation}{0}

{\large{1.~Introduction}}

\vspace{5mm}

There has been growing interest in the response of semiconductor
superlattices to high--frequency ac electric
fields~\cite{TSU,IGN,REV}.
 The first studies of superlattices in far--infrared laser fields

were reported~\cite{GUI} very recently.
It is now possible to investigate experimentally many of the
theoretical predictions concerning photon--assisted tunneling, or the
influence of an external terahertz
electric field on miniband transport.

The obvious first approximation for a theoretical description of
semiconductor superlattices is an ideal perfectly periodic structure,
which permits the usual simplifications implied by the Bloch theorem.
This is equally true even in the presence of a spatially homogeneous
time dependent field of arbitrary strength~\cite{QWS}.  But, of
course, actual systems of mesoscopic size inevitably contain
imperfections.    First of all, the actual number of

superlattice periods is finite, and relatively small. In addition,
since fabrication

is always somewhat imperfect, the individual quantum wells inevitably
differ

from one another in size, shape, and separation.   In the more
familiar situation of {\em{equilibrium}} many particle systems the
special
case of local perturbations --- even strong ones --- is amenable to
ready
analysis through the use of standard Green's function techniques, as
long as
one has available the solution to the ideal unperturbed problem. Our
goal
here is to extend these techniques for handling local perturbations
to the
realm of systems under the influence of strong time periodic fields.

In the following section we will develop a formal
approach to this problem. In  section~3 we illustrate the results for
several
different kinds of defects by numerical model calculations. The paper
concludes
with a discussion.

\vspace{1cm}

\setcounter{chapter}{2}
\setcounter{equation}{0}

{\large{2.~Green's functions for lattices in external ac fields}}

\vspace{5mm}

Before turning to systems in time periodic fields, we
review briefly the origin of the simplicity of spatially localized
perturbations
in equilibrium systems. Then we will formulate the problem of
interest so that it can be treated by an analogous formalism.

For a system described by a time independent Hamiltonian $H = H_{0} +
V $
we consider the resolvent operator $ R(z) $ as a function of the
complex
frequency $z$:
\be
R(z) = \frac{1}{z - H} = R^{0}(z) + R^{0}(z) V R(z) \;\;\; ,
\label{LSE}
\ee
where $ R^{0}(z) = 1/(z - H_{0}) $ is the unperturbed resolvent
operator.
The singularities of $ R(z) $ as a function of $z$ are at the energy

eigenvalues of the Hamiltonian, so it contains the fundamental
spectral
information which controls the system dynamics. As is particularly
convenient
in the study of disordered many particle systems, the resolvent
operator can be

replaced by the single particle Green's function, essentially a
matrix element
of that operator; the Lippmann--Schwinger equation~(\ref{LSE})
becomes the
Dyson equation, of the same structural form. In either case, if the
perturbation
$ V $ has nonvanishing matrix elements only between a finite number
of
spatially localized states in a complete set, then the equation is
readily
solved by finite quadratures, in terms of the unperturbed solutions $
R^{0} $
(or the corresponding Green's functions). In that representation of
spatially
localized states we have
\be
R_{j\ell} = R^{0}_{j\ell} + \sum_{j'\ell'} R^{0}_{jj'} V_{j'\ell'}
R_{\ell'\ell}

\;\;\; .        \label{DYS}
\ee
Because of the locality of $ V $ the right hand side involves $
R_{\ell'\ell} $
only for a small number of sites $\ell'$. Therefore,
after~(\ref{DYS}) has been
inverted for that subset of resolvent operator matrix elements, the
equation
itself gives $ R_{j\ell} $ for {\em{all}} sites $j$.

The first step in developing a corresponding formalism for electrons
in an imperfect lattice

interacting with an external time--periodic field is to cast the
Schr\"{o}dinger
equation for that case into the form of a stationary eigenvalue
problem. For a system with a periodically
time dependent Hamiltonian, $ H(t) = H(t + T) $, the Floquet
theorem~\cite{FLO}
guarantees that  the Schr\"{o}dinger equation (we choose units such
that $\hbar=c=1$),
\be
 \left( H(t) - i\partial_{t}\right) \psi(t) = 0 ,
\ee
has a complete set of solutions that are of the special form
\be
\psi_{\alpha}(t) = u_{\alpha}(t) \exp(-i\varepsilon_{\alpha}t)

\label{FLO}
\ee
with functions $ u_{\alpha}(t) $ that inherit the periodicity of the

Hamiltonian:
\be
u_{\alpha}(t) = u_{\alpha}(t+T) \;\;\; .
\ee
These ``Floquet functions'' $ u_{\alpha}(t) $ then obey the
eigenvalue

equation
\be
\left( H(t) - i\partial_{t} \right) u_{\alpha}(t) =

\varepsilon_{\alpha}u_{\alpha}(t)       \label{EIG}
\ee
with eigenvalues $ \varepsilon_{\alpha} $, the ``quasienergies'', and
eigenoperator
\be
{\cal{L}}(t) \equiv H(t) - i\partial_{t}        \;\;\; .
\ee
In order to extend the standard formalism of linear vector spaces in
quantum

mechanics to these problems, it is necessary to introduce~\cite{SAM}
an extended Hilbert
space of time--periodic functions with the scalar product
\be
\langle\!\langle u_{1}(t) | u_{2}(t) \rangle\!\rangle =
\frac{1}{T}\int_{0}^{T} \! dt
\, \langle u_{1}(t) | u_{2}(t) \rangle     \;\;\; ,
\ee
where the single angle brackets denote the usual scalar product. An
essential
complication of the time--periodic problem appears here. If $
u_{\alpha}(t) $
is a Floquet function with quasienergy $ \varepsilon_{\alpha}$ then,
for any
(positive or negative) integer $n$, the product
$u_{\alpha}(t) \exp(in\omega t) $, where $\omega\equiv 2\pi/T$,

is also a Floquet function, with quasienergy

$ \varepsilon_{\alpha} + n\omega $.

These functions are all equivalent in terms of the full solution
$ \psi_{\alpha}(t) = u_{\alpha}(t)\exp(-i\varepsilon_{\alpha}t) $
of the Schr\"{o}dinger equation, simply redefining the separation
into periodic
Floquet function and quasienergy phase factor, but they are
{\em{not}} equivalent

for the related eigenvalue formalism associated with Eq.~(\ref{EIG}).
In particular,

they are all needed for the completeness relation in the extended
Hilbert space, which covers only the restricted time interval
$[0,T]$.
As we shall see, this fact greatly complicates the solution of the
formal equivalent
of the Dyson equation~(\ref{DYS}) for a lattice with a local
perturbation.

Even though Floquet theory is, in principle, applicable to problems
where the

external ac field induces strong transitions between different energy

bands~\cite{ACS}, we will  limit ourselves here, for simplicity,  to
situations where the probabilities

of such interband transitions are negligibly small, so that is
suffices to treat the

dynamics within a single band. The spatial periodicity of the ideal
lattice gives

Bloch wave solutions to the Schr\"{o}dinger equation in the absence
of the

time--periodic driving force:
\be
\varphi_{k}(x) = e^{ikx} v_{k}(x)  \;\;\; ;        \label{BLO}
\ee
the function $ v_{k}(x) = v_{k}(x + a) $ is spatially periodic with
the lattice
period $a$. We denote the corresponding energy eigenvalues by $ E(k)
$.
If we introduce a spatially uniform, time--dependent electric field,

using a transverse gauge:
$ {\cal{E}}(t) = -d A(t) / dt $, with the electromagnetic vector
potential
$ A(t) $, then we can include its effect in the Hamiltonian by the
usual
replacement $ p \rightarrow p - eA(t) $. It follows that the solution
to the

time--dependent Schr\"{o}dinger equation is given
by~\cite{QWS,HOU,KRI}
\be
\psi_{k}(x,t) = e^{ikx} v_{q(t)}(x) \exp\left\{ -i\int_{0}^{t} \!
d\tau \,
E[q(\tau)] \right\}      \;\;\; ,        \label{WAV}
\ee
where $ v_{k}(x) $ on the right hand side is the solution~(\ref{BLO})
to the
field--free problem, and we have introduced the notation
\be
q(t) = k - eA(t)        \;\;\; .
\ee
If $A(t)$ is periodic, we can write the wave functions~(\ref{WAV}) in
the standard Floquet
form~(\ref{FLO}), with quasienergies
\be
\varepsilon(k) = \frac{1}{T}\int_{0}^{T} \! d\tau \, E[q(\tau)]

\ee
and time--periodic Floquet functions
\be
u_{k}(x,t) = e^{ikx} v_{q(t)}(x) \exp\left\{ -i\int_{0}^{t} \! d\tau
        (E[q(\tau)] - \varepsilon(k) ) \right\}     \;\;\; ,
\ee
so that
\be
\psi_{k}(x,t) = u_{k}(x,t) \exp(-i\varepsilon(k)t)      \;\;\; .
\ee
In the spirit of the Floquet approach, we also need all the
``satellites''
$ u_{k}(x,t) \exp(in\omega t) $, which we denote by $ | k,n\rangle $,
so that $ | k,0 \rangle $ is the representative that evolves into the
unperturbed

Bloch state labeled by $k$ when the amplitude of the external field
vanishes.

Correspondingly, if $ \{ | \ell \rangle \} $ is a  set of time
independent localized  states,
complete in the usual Hilbert space,
we use the abbreviation $ | \ell, m \rangle $ for the states

$ | \ell \rangle \exp(im\omega t) $. Thus, if we again introduce the

perturbation $V$ (which, in  light of the motivating examples in the
introduction, we take to be time--independent) into the Hamiltonian,
so that
$ {\cal{L}} = {\cal{L}}_{0} + V $, then we can define the Green's
functions
\begin{eqnarray}
G^{0}_{j\ell}(n,m) & \equiv & \langle\!\langle j, n | 1 / (z -
{\cal{L}}_{0})
        | \ell, m \rangle\!\rangle  \label{DFG} \\
G_{j\ell}(n,m) & \equiv & \langle\!\langle j, n | 1 / (z - {\cal{L}})
        | \ell, m \rangle\!\rangle        \;\;\; ,
\end{eqnarray}
whose singularities now occur at the quasienergies, again
central to the description of the system dynamics.
(We have suppressed the explicit dependence of the Green's functions
on $z$ so as not to clutter the notation). The Dyson equation becomes
\be
G_{j\ell}(n,m)= G^{0}_{j\ell}(n,m) + \sum_{j',\ell',n'}
G^{0}_{jj'}(n,n')

        V_{j'\ell'} G_{\ell'\ell}(n',m)     \;\;\; .
\label{EDS}
\ee
Even if the locality of $ V $ restricts the values of the indices $
j' $ and

$ \ell' $, as in the time--independent case~(\ref{DYS}), there
remains
an infinite sum over the ``photon index'' $ n' $ which labels
different

representatives of the Floquet functions. For each pair of site
indices
$ j,\ell $, the Green's functions $G_{j\ell}(n,m)$ are infinite
matrices in the photon index space.

We take the spatially localized states labelled above by the lattice
site
indices $ \ell $ to be the Wannier states appropriate to the band of
interest.
Then, within the standard tight binding approximation ($v_k(x)$ in
Eq. (\ref{BLO}) independent of wave vector $k$), we have
\be
\langle \ell | k \rangle = e^{ik\ell a}/\sqrt{N} \;\;\; ,
\ee
where $N$ is the number of wells. After inserting a complete set of
Floquet functions $ \{ | k,s \rangle \} $ into
the definition~(\ref{DFG}), it follows that the unperturbed Green's
functions

are given by
\begin{eqnarray}
G^{0}_{j\ell}(n,m) & = & \sum_{k,s} \langle\!\langle j,n | k,s
\rangle\!\rangle
        \frac{1}{z - \varepsilon(k) - s\omega}
        \langle\!\langle k, s | \ell,m \rangle\!\rangle    \nonumber
\\
& = & {1\over N}\sum_{k,s}  \frac{e^{ika(j - \ell)}}{z -
\varepsilon(k) - s\omega}
        F_{s-n}(k) F^{*}_{s-m}(k)       \;\;\; ,        \label{GRE}
\end{eqnarray}
where the quantities $ F_{n}(k) $ are the Fourier coefficients of the
time periodic function
\be
F(k,t) \equiv \exp\left\{ -i\int_{0}^{t} \! d\tau \,

        (E[q(\tau)] - \varepsilon(k) ) \right\}

        = \sum_{n} F_{n}(k) e^{-in\omega t}     \;\;\; .
\label{FKT}
\ee
If, for a given energy dispersion $ E(k) $,  these Fourier
coefficients
decay sufficiently rapidly, so that only a small number of terms is
significantly

different from zero, then the summation over the satellites $s$
in~(\ref{GRE})

can be restricted to a small set. If, for consistency, we terminate
the Fourier expansion of the full perturbed Floquet states in the
same way, then the Dyson equation~(\ref{EDS})

becomes an equation for finite--dimensional matrices.

If we can neglect all but a single satellite, we can return to the
simplicity
of the time--independent problem.  An important example is the case
where the band
width $ W $ is significantly smaller than the photon energy $ \omega
$,
while $ \omega $ is itself small compared to the separation between
different bands, so that interband transitions remain negligible.
The crude estimate~\cite{fnote}
\be
\int_{0}^{t} \! d\tau \, ( E[q(\tau)] - \varepsilon(k) ) < WT
        = 2\pi W/\omega \;\;\; , \label{LWT}
\ee
valid for {\it all} times $t$, shows that the exponent in~(\ref{FKT})
is, at most, of  order $ W/\omega $.
In the high frequency limit $ \omega \gg W $ we retain

only the leading term: $ F_{0}(k) \approx 1 $,
so that $ G^{0}_{jl}(n,m) $ becomes diagonal in the photon indices:
\be
G^{0}_{jl}(n,m) \approx \frac{1}{N}\sum_{k} \frac{e^{ika(j-l)}}{z -
\varepsilon(k) - n\omega}\delta_{nm}    \;\;\; .
\label{UGF}
\ee
Within this approximation the full propagator $G$ is also diagonal in
the photon indices, and Eq.~(\ref{EDS}) takes the simple form of
Eq.~(\ref{DYS}); each Green's function is labelled simply by two site
indices $ j $ and $ \ell $.  (The equation is of the same structure
for any
value of the photon index $n$, and the functions for different values
of $n$ are independent. The only impact of that index is to shift the
value of $z$ by $n\omega$ everywhere.  For simplicity we will always
take $n=0$ in the expressions below.)  These propagators are of the
standard analytic
form: they have poles at each of the quasienergies, $ z =
\varepsilon(k) $

(or, in the continuum limit, a cut on the real $z$--axis throughout
the

quasienergy band).

As in the time--independent case, {\em{spectral}} isolation of a
quasienergy
eigenvalue implies {\em{spatial}} localization of the corresponding
Floquet
state in the neighborhood of the defect. We can also determine the
time
averaged spatial distribution of states associated with such isolated
poles
of the resolvent operator. Following the usual many body,
nonrelativistic
field theory formalism, we perform a formal spectral decomposition of
the
site diagonal Green's functions, using the eigenstates of the full
dynamical
operator $ {\cal{L}} = H(t) - i\partial_{t} $, including the defect:
\be
{\cal{L}} | \alpha, m \rangle =
(\varepsilon_{\alpha} + m\omega) | \alpha, m \rangle \;\;\; .
\ee
The site diagonal Green's function at complex quasifrequency $z$ is
\be
G_{\ell\ell}(n,n) = \langle\!\langle \ell, n | 1 / (z - {\cal{L}}) |
        \ell, n \rangle\!\rangle = \sum_{\alpha, m}
        \frac{ | \langle\!\langle \ell, n | \alpha, m
\rangle\!\rangle |^{2} }
        {z - \varepsilon_{\alpha} - m\omega }   \;\;\; .
\ee
Now suppose that $ z = \varepsilon_{\beta} $ is an isolated pole
($\varepsilon_{\beta}\neq \varepsilon_{\alpha} + m\omega$ for any
integer $m$

and for any $\alpha\neq\beta$).
If we  sum  all the residues of its satellites,
\be
\sum_{m} {\mbox{Res}} \,
\left. G_{\ell\ell}(n,n) \right|_{z = \varepsilon_{\beta} + m\omega}
= \sum_{m} | \langle\!\langle \ell, n | \beta, m \rangle\!\rangle
|^{2} \;\;\; ,
\ee
and use the equivalence of the sum over $m$ to that over $n$, then
with Parseval's identity,
\be
\sum_{n} \langle\!\langle \psi | \ell, n \rangle\!\rangle
        \langle\!\langle \ell, n | \phi \rangle\!\rangle
        = \frac{1}{T} \int_{0}^{T} \! dt \, \langle \psi(t) | \ell
\rangle
        \langle \ell | \phi(t) \rangle  \;\;\; ,
\ee
we obtain
\be
\sum_{m} {\mbox{Res}} \,
\left. G_{\ell\ell}(n,n) \right|_{z = \varepsilon_{\beta} + m\omega}
= \frac{1}{T} \int_{0}^{T} \! dt \, | \langle \ell | \beta(t) \rangle
|^{2}
\;\;\; .        \label{RES}
\ee
The sum over the residues of all satellites of an isolated pole

$ \varepsilon_{\beta} $ is therefore just the probability for the
Floquet
state $ | \beta(t) \rangle $ to lie in the $\ell$--th quantum well,
time
averaged over the period~$T$. For later use we remark that in the

high--frequency regime, where the function $ F(k,t) $

is approximated by unity, the sum in~(\ref{RES}) is saturated by the

single term with $m = 0$.  This result then has the familiar
structure of the corresponding time independent problem.

\vspace{1cm}

\setcounter{chapter}{3}
\setcounter{equation}{0}

{\large{3.~Examples}}

\vspace{5mm}

In order to apply the general theory we now introduce a specific
model
Hamiltonian. Corresponding electronic levels within the individual
quantum wells of a semiconductor superlattice

are weakly coupled by quantum tunneling. As long as the dynamics in
the

plane of the wells are decoupled from the one--dimensional dynamics
along the
lattice direction, the electronic behavior associated with a single
such set of tunneling-coupled levels is well described  by a standard

tight binding Hamiltonian:
\be
H_{tb} = E_{0} \sum_{\ell} | \ell \rangle \langle \ell |

+ \frac{W}{4} \sum_{\ell} \left( | \ell \rangle \langle \ell + 1 | +
        | \ell + 1 \rangle \langle \ell | \right)  \;\;\; ,
\label{HTB}
\ee
where $ \ell $ labels the wells, and $ W $ is the width of the
resulting
energy band. This Hamiltonian gives the familiar dispersion relation
\be
E(k) = E_{0} + \frac{W}{2} \cos(ka)     \;\;\; .        \label{DIR}
\ee
We choose the zero of energy to be at the band center: $ E_{0} = 0 $.
We assume that an external laser field is linearly polarized in the
superlattice direction, so that a one--dimensional description
remains
possible. For a homogeneous electric field with amplitude $
{\cal{E}}_{0} $
and frequency $ \omega $,
\be
{\cal{E}}(t) = {\cal{E}}_{0} \sin(\omega t)     \;\;\; ,
\ee
we find for the quasienergies
\begin{eqnarray}
\varepsilon(k) & = &

\frac{W}{2} J_{0} \! \left(\frac{e{\cal{E}}_{0}a}{\omega}\right)
\cos(ka)
\;\;\; \bmod \omega     \nonumber \\
& \equiv & C \cos(ka) \;\;\; \bmod \omega     \;\;\; ,  \label{QES}
\end{eqnarray}
where $ J_{0}(x) $ is the zero order Bessel function.
The parameter $ C $ that we have introduced here is the half width of
the

quasienergy band.  Note that, by definition, $C$ can become equal
to zero (when $ e{\cal{E}}_{0}a/\omega $ is equal to a zero of
$J_{0}$),
or even negative.

We will now  be concerned solely with the high--frequency regime $
\omega > W$.
The unperturbed Green's functions~(\ref{UGF}) are given by
\be
G^{0}_{j\ell} \; = \; \frac{2}{C} \frac{\zeta_{0}^{|j-\ell| + 1}}{1 -
\zeta_{0}^{2}}
\; \equiv \; g(|j - \ell|)    \label{IGF}
\ee
with
\be
\zeta_{0} = \frac{z}{C} - \sqrt{(z/C)^{2} - 1}          \;\;\; .
\label{POL}
\ee
Mathematically, these are poles of the integrand in~(\ref{UGF}),
which defines

$ G^{0} $ as an integral around the unit circle in the complex plane
in the variable $e^{ika}$.   For $z$ real and $|z/C|>1$ the sign of

the square root is to be taken so that they are the poles {\it
within} the unit circle.  For values of
$ z $ real and within the quasienergy band, the radicand
in~(\ref{POL})
is negative, and $ G^{0} $ is complex (the corresponding branch cut
of
$ G^{0}(z) $ in the complex $z$--plane, reflected by the change in
sign
of the square root in~(\ref{POL}), is on the real axis, along the
range of
quasienergy eigenvalues).

As we have noted in section~1, a particularly important example of a
local
perturbation is the termination of the chain of quantum wells at both
ends.

The most trivial effects of this, the cutting of the infinite chain
to eliminate

hopping beyond the end wells, are accommodated by introducing
suitable

boundary conditions on the wave functions:
$ \sqrt{N}\langle \ell | k \rangle = \sqrt{2}\sin(k\ell a) $, rather
than $ e^{ik\ell a} $,
with $ k(N+1)a = r\pi $ for a chain of $ N $ wells; $ r = 1,2, \ldots
, N $.
The quasienergy eigenvalues remain unchanged from~(\ref{QES}),
except that $ k $ is restricted to discrete values. If $ N+1 \approx
N $ is
sufficiently large that we can, nonetheless, approximate the sum over
$k$
in~(\ref{UGF}) by an integral, then the Green's functions become
\be
G^{1}_{j\ell} = {2\over N}\sum_{k} \frac{ \sin(k j a) \sin(k\ell
a)}{z - \varepsilon(k)}
\approx g(|j - \ell|) - g(j + \ell)         \;\;\; .  \label{FGF}
\ee
Of course, these Green's functions no longer only depend on the well
separation $ | j - \ell | $ (the argument $ (j + \ell) $
in~(\ref{FGF}) has to be
taken modulo $N$). But there are additional inevitable modifications
to the Hamiltonian of a finite chain. First, the end wells are, in
general,
different from the others; termination might, e.g., correspond to a
very high
potential barrier at the chain edge. The energy levels in those wells
will
then be different, by an amount $ \nu $, from that in all the other
wells.
If we label the wells from $1$ to $N$, then the perturbation is
$ V_{11} = V_{NN} = \nu $, and the Dyson equation~(\ref{EDS}) reads
\be
G_{j\ell} = G^{1}_{j\ell} + \nu \left(G^{1}_{j1}G_{1\ell}
        + G^{1}_{jN}G_{N\ell} \right)    \;\;\; .
\ee
The two coupled equations for $ j = 1 $ and $ j = N $ give
\be
G_{1\ell} = \frac{(1 - \nu G^{1}_{11}) G^{1}_{1\ell}

        + \nu G^{1}_{1N} G^{1}_{N\ell} }
        {(1 - \nu G^{1}_{11})^{2} - \nu^{2} (G^{1}_{1N})^{2} }
        \approx \frac{G^{1}_{1\ell}}{1 - \nu G^{1}_{11}}
\;\;\; ,
\label{FIN}
\ee
and from the general results~(\ref{IGF}) and (\ref{FGF}) we obtain
\be
G^{1}_{1\ell} = G^{1}_{N,N-\ell} = \frac{2\zeta_{0}^{\ell}}{C}
\;\;\; .
\label{G1L}
\ee
The final approximate form in~(\ref{FIN}) neglects the propagator

$ G^{1}_{1N} $ between the ends of the chain, which by~(\ref{G1L}),
and for

obvious physical reasons, vanishes with increasing chain length. If
the

denominator in~(\ref{FIN}) vanishes ($G_{1\ell}$ has a pole) on the
real axis

beyond the branch cut $ |z| < C $ (note from the
definition~(\ref{POL}) that on the
real axis $ | \zeta_{0} | < 1 $), then this value of $z$ corresponds
to an eigensolution
spatially localized near the chain end, with quasienergy eigenvalue
\be
\varepsilon = \frac{C}{2} \left( \zeta_{0} +
\frac{1}{\zeta_{0}}\right)
\approx \nu + \frac{C^{2}}{4\nu}        \;\;\; .        \label{LOM}
\ee
The requirement that the corresponding value of $ | \zeta_{0} | $,
which is $ | C / 2\nu | $, be less than unity leads to the condition
\be
| J_{0}(e{\cal{E}}_{0}a/\omega) | < \frac{4 | \nu |}{W}  \;\;\; .
\label{INE}
\ee
Thus, we see the possibility of using the laser parameters to control
the appearance (or disappearance) of local modes: If $ 4 | \nu | /W <
1 $,
the inequality~(\ref{INE}) can not be satisfied for small amplitudes
$ {\cal{E}}_{0} $. The localized solution can only exist if the
argument
$ e{\cal{E}}_{0}a/\omega $ makes the Bessel function sufficiently
small.

If we keep the additional terms in the denominator of~(\ref{FIN})
which couple the two ends of the chain, then the isolated pole
splits into two (the even and odd parity combinations with respect
to the chain center, as usual); the splitting is approximately
given by
\be
\Delta \varepsilon = C \left( \frac{C}{2\nu} \right)^{N-2}\left[1-

\left({C\over 2\nu}\right)^2\right]       \;\;\; .
\label{SPL}
\ee

To illustrate the theory with a numerical example, we employ a model
potential which consists of a finite array of square wells with
typical
superlattice parameters: the well width is 100~\AA, the barrier width
50~\AA,
and the barrier height 300~meV. The particle moving in this potential
is
taken to have a mass of 0.067 relative to a bare electron (this is
the effective mass in the conduction band of GaAs).
For these parameters, the lowest band has a width $ W = 1.813$~meV;
it is separated from the next band by a gap $E_g = 92.5$~meV. The
next--to--nearest
neighbor hopping elements are two orders of magnitude smaller than
the

nearest neighbor couplings $ W/4 $, so that the single band tight
binding

Hamiltonian~(\ref{HTB}) provides a fairly good description of each
band of the system.

Fig.~1 shows the lowest quasienergy band for this model with $ N = 50
$
wells and a driving frequency $ \omega = 10.0 $ meV, versus the
amplitude
${\cal{E}}_{0}$ of the ac field.  Since $\omega\ll E_g$, interband
effects can be neglected as long as $e{\cal E}_0a\ll E_g$ (for
sufficiently short times; ultimately, of course, there is a
substantial probability for a real interband transition).  It turns
out that the change $\nu$ of the on--site

energy at the ends of the chain is almost equal to $ W/4 $, so that
the local
mode~(\ref{LOM}) should appear as soon as $ {\cal{E}}_{0} > 0 $.
Indeed,
eq.~(\ref{LOM}) describes the isolated quasienergies on top of the
``miniband''
in Fig.~1 quite well, as expected for a ratio $ \omega/W = 5.5 $.

The splitting $ \Delta\varepsilon $ of the local modes is too small
to be seen
in Fig.~1, but from~(\ref{SPL}) it is clear that $ \Delta\varepsilon
$ will

become larger when the number of wells $N$ is decreased. Fig.~2
confirms
this for a lattice with only 10 wells; the other parameters are the
same as

those employed in the previous figure.

The theory can be refined further. For example, as the wave function
in the
initial well is changed by the boundary at the chain end, so must be
its

overlap with, and therefore its hopping rate to, the state in the
neighboring

well. The potential $V$ then contains the additional terms
$ V_{12} = V_{21} = V_{N-1,N} = V_{N,N-1} \equiv \delta $.
As we have already seen, it is somewhat simpler algebraically to look
at
a single termination, i.e.~a semi--infinite chain, and for a
reasonably long

finite chain the ends act nearly independently, so we give those
results
explicitly. The general Dyson equation becomes
\be
G_{j\ell} = G^{1}_{j\ell} + \nu G^{1}_{j1} G_{1\ell} + \delta

\left[ G^{1}_{j1} G_{2\ell} + G^{1}_{j2} G_{1\ell} \right]
\;\;\; .
\ee
Again this equation couples only a single pair, $ j = 1 $ and $ j = 2
$,
with the result
\be
G_{1\ell} = \frac{(1 - \delta G^{1}_{12}) G^{1}_{1\ell}
        + \delta G^{1}_{11}G^{1}_{2\ell} }
{1 - \nu G^{1}_{11} - 2 \delta G^{1}_{12}
        + \delta^{2} [ (G_{12}^{1})^{2} - G^{1}_{11} G^{1}_{22} ] }
\;\;\; .
\label{HOP}
\ee
To obtain an explicit expression for the isolated poles we need, in
addition to the unperturbed Green's functions~(\ref{G1L}), only
one more: $ G^{1}_{22} = 2\zeta_{0}(1 + \zeta_{0}^{2}) / C $.
Then the zeros in the denominator of~(\ref{HOP}) are found at
\be
\zeta_{0} = \frac{-C}{4\delta(C+\delta)} \left[ \nu \pm
\sqrt{\nu^{2} + 4\delta(C + \delta)} \right]    \;\;\; .
\label{PVD}
\ee
For sufficiently small $ | C | $ (sufficiently close to the zeros of
$ J_{0} $,
at the points of the band collapse, so that $ | \zeta_{0} | < 1$)
both solutions exist. The corresponding quasienergies are
\be
\varepsilon = \frac{1}{2} \left[ \nu \mp
\sqrt{\nu^{2} + 4\delta(C + \delta)} \right]
-\frac{C^{2}}{8\delta(C + \delta)}
\left[\nu \pm \sqrt{\nu^{2} + 4\delta(C + \delta)} \right]
\;\;\; .
\ee
However, for a realistic lattice the change $ \delta $ of hopping
strength

at the chain ends is of the same order of magnitude as the
next--to--nearest

neighbor couplings. In order to describe these effects, a term
proportional to
$ \cos(2ka) $ has to be added to the dispersion~(\ref{DIR}), and the
calculation of the unperturbed Green's functions then involves the
solution
of a fourth order equation.

More important types of defects in semiconductor superlattices are
the
inevitable irregularities in the width of the barriers or wells.
For algebraic simplicity we treat the limit of long chains,
appropriate to

defects far from the ends of the physical lattice. First we consider
a system

with a  barrier defect:  we assume that the width of the barrier
between
the wells labelled, for convenience, by 0 and 1  is different from
the others.
This simplest perturbation has no effect on the on--site energies in
the tight

binding Hamiltonian~(\ref{HTB}), but alters only the hopping between
the wells
separated by this barrier:
\be
V_{j\ell} = \Delta(\delta_{j 0}\delta_{\ell 1} +
\delta_{j1}\delta_{\ell 0} )
\;\;\; .
\ee
The Dyson equation~(\ref{EDS}),
\be
G_{j\ell} = G^{0}_{j\ell} + \Delta\left(
        G^{0}_{j0}G_{1\ell} + G^{0}_{j1}G_{0\ell} \right) \;\;\; ,
\ee
couples directly only $ j = 0 $ and $ j = 1 $, to give
\be
G_{0\ell} = \frac{ [ 1 - \Delta G^{0}_{10} ] G^{0}_{0\ell}
        + \Delta G^{0}_{00}G^{0}_{1\ell} }
{ (1 - \Delta G^{0}_{10} )^{2} - \Delta^{2} ( G^{0}_{00} )^{2} }
\;\;\; .
\ee
With the explicit expression~(\ref{IGF}) for $ G^{0}_{j\ell} $ we
then find
the potential isolated poles to be at
\be
\zeta_{0} = \pm \frac{C}{C + 2\Delta}   \;\;\; ;        \label{BAR}
\ee
the quasienergies are
\be
\varepsilon = \pm \left( C + \frac{2\Delta^{2}}{C + 2\Delta} \right)
\;\;\; .
\ee
Since we require $ | \zeta_{0} | < 1 $, the solution can only exist
if
either $ \Delta/C > 0 $ or $ \Delta/C < -1 $. Where the solution does
exist, the poles occur at pairs of quasienergies of equal magnitude
and opposite sign; at the points of the band collapse, where $ C = 0
$,
these are at $ \varepsilon = \pm \Delta $.

At this point, it is of interest to recall that $ C $ changes its
sign when
$ e{\cal{E}}_{0}a/\omega $ is equal to a zero of $ J_{0} $. Let us
assume
that the defect is a barrier  shorter than the other ones, so that $
\Delta $
is positive. Let us keep $ \omega $ fixed, and let $
{\cal{E}}_{0}^{(1)} $
and $ {\cal{E}}_{0}^{(2)} $ be the field amplitudes that correspond
to the first
two zeros of $ J_{0} $. Then according to~(\ref{BAR}) the localized
solutions
will exist in the full interval $ 0 \leq {\cal{E}}_{0} \leq
{\cal{E}}_{0}^{(1)} $,
but for amplitudes between $ {\cal{E}}_{0}^{(1)} $ and $
{\cal{E}}_{0}^{(2)} $
they should appear only where $ | C | < \Delta $.

Fig.~3 shows an example of a quasienergy band for a system with a
barrier defect. The same model potential as before has been used with
$ N = 50 $ wells. Near the middle of the chain one barrier is 10
\AA~shorter
than the other ones, which increases the hopping strength by
$ \Delta = 0.421$~meV, to be compared to the full band width of
$ W = 1.813$~meV. The frequency is $ \omega = 5.0$~meV.
In addition to the edge states discussed before, Fig.~3 shows the
quasienergies
of the two states associated with the defect, symmetrically above
and below the band. But these states behave differently from what the
high--frequency approximation~(\ref{BAR}) predicts: although they
join the
band immediately after the first collapse, they reappear when the
amplitude
is only slightly increased. To obtain a correct analytical
description of this

case, it is apparently not sufficient to simply approximate
(\ref{FKT}) by unity.

We might also look at the case where a single well differs in width
from all the others. In the simplest approximation this only changes
the energy level in this well by an amount~$v$:
$ V_{j\ell} = v \delta_{j0} \delta_{\ell 0} $. The propagators are
then
\be
G_{j\ell} = G^{0}_{j\ell} + \frac{v G^{0}_{j0}G^{0}_{0\ell}}
{ 1 - v G^{0}_{00} }  \;\;\; ;  \label{PDW}
\ee
the isolated quasienergies are given by
\be
\varepsilon = {\mbox{sign}}(v) \sqrt{ v^{2} + C^{2} }       \;\;\; .
\label{TOP}
\ee

In Fig.~4 we display the quasienergies for our model system,
where now one of the wells has been shortened from 100~\AA~to
98~\AA, resulting in $ v = 1.084$~meV. The approximation~(\ref{TOP})
yields a good description of the defect state on top of the band.

Finally, we use this example to study the spatial localization of
defect states.  As already stated, within the high--frequency
approximation
the sum over the satellites in~(\ref{RES}) is saturated by a single
term.
{}From~(\ref{PDW}) the residue of $ G_{\ell\ell} $, which is the
probability of
finding the state supported by the defect in the $\ell$--th well, is
readily

found to be
\be
\frac{| v |}{\sqrt{v^{2} + C^{2}}} \left[
\sqrt{\left(\frac{v}{C}\right)^{2}
        + 1} - \frac{|v|}{C} \right]^{2|\ell|}     \;\;\; .
\ee
This falls off exponentially with the distance $ {\ell} $ from the
defect,

which was chosen to be on the site labelled 0. The decay rate,
$ -2\ln \left[\sqrt{(v/C)^{2} + 1} - |v|/C\right] $,
increases monotonically with $ |v|/C $, the ratio of the size of the

perturbation relative to the half width of the quasienergy band of
the perfect
system at the laser parameters of interest. In particular, at the
points of
the band collapse the local state should be concentrated wholly in
the

anomalous well.

To check this prediction, we introduce a rather small defect into our
model potential and shorten one well by 0.5~\AA, which gives

$ v = 0.266$~meV. Fig.~5a shows the probability density of the
Floquet
state associated with the defect at $ \omega = 5.0$~meV and
$ {\cal{E}}_{0} = 3000$~V/cm, where it is still delocalized over
several
wells in the neighborhood of the defect. But as shown in Fig.~5b,
at $ {\cal{E}}_{0} = 8016.08$~V/cm, right at the first collapse, the
state
is completely contained within the single narrower well itself.

\vspace{1cm}

\setcounter{chapter}{4}
\setcounter{equation}{0}

{\large{4.~Discussion}}

\vspace{5mm}

We have shown how to extend the powerful Green's function formalism
from time independent equilibrium systems to electrons interacting
with strong external homogeneous ac electric fields, by taking
advantage of the special simplifications of temporal periodicity.
In particular, with the introduction of quasienergy bands for
independent electrons in spatially periodic lattices we have been
able to treat analytically the combined
effect of spatially localized perturbations and the strong time
dependent external field.  However, some of the simplicity of the
analogous time--independent
problem is lost because of the periodically repeated  structure of
the quasienergy
spectrum (period $\omega$): even for a localized defect, the Dyson
equation~(\ref{EDS}) contains
an infinite summation over the ``photon index'' $n'$.

For  practical calculations this infinite sum must be truncated. In
the
high--frequency limit of a single band system, where the ac frequency
$ \omega $ is significantly larger than the band width $ W $, the sum
can
approximately be replaced by a single term. In this approximation the
effect
of the ac field is completely described by the parameter $ C $, the
half width
of the quasienergy band at a given frequency and amplitude of the
external field.  Even this
simple approximation leads to nontrivial predictions, such as the
possibility
of controlling the appearance, or
disappearance, of localized states by adjusting the parameters of the
ac field. Such a differential tuning of local and
extended states is not entirely novel: magnetic impurities with
Land\'{e}
$g$--factors suitably different from those of the host, for example,
or in antiferromagnets, can lead~\cite{MAG} to localized
magnon (spin wave) modes which can be tuned into or out of the magnon
continuum with static magnetic fields.

However, when the frequency becomes smaller than the band width, the
quasienergy
spectrum --- even of a single band system --- will always be the
whole real
axis, and a quasienergy eigenvalue of a defect state, split off from
the original

band, will always be embedded in the continuum of a satellite band.
In such a case, a systematic expansion of the function $ F(k,t) $

(see~(\ref{FKT})) becomes necessary. Even though working with the
Dyson

equation~(\ref{EDS}) may then be quite cumbersome, it is, in
principle, possible
to keep track systematically  of defect--induced multiphoton
processes.

The framework outlined in the present paper is not limited to the
case of
a single band. In fact, the field--induced interaction of different
bands can have very interesting effects. We will give two numerical
examples.
First, we consider a dimerized system (analogous to a Peierls
unstable atomic chain, such as polyacetylene~\cite{SSH}), where the
barrier width is 40~\AA~throughout
a chain of 50 wells, but the well width alternates between 90 and
110~\AA.
As before, the barrier height is $0.3$~eV, and the effective particle
mass
is 0.067 electron masses. The bands of such a dimerized system appear
in closely spaced doublets, with a spacing that can be tuned (in
contrast to the atomic chain case) by varying the

difference between the widths of the two types of wells.

Fig.~6 shows the lowest doublet of quasienergy bands for

$ \omega = 2.0$~meV; the only ``defect'' is the finite size of the
system.

For amplitudes lower than $ {\cal{E}}_{0} = 5000$~V/cm the two
members of
the doublet hardly influence each other and show the Bessel
function--type

collapse pattern, as expected for single band systems with a total
period of

$ a = 280$~\AA. But for larger amplitudes the behavior changes due to

interactions between the bands. At 10~kV/cm, the quasienergy bands
are

even wider than the original unperturbed energy bands, and additional
edge

states emerge.

Another example is shown

in Fig.~7.  In this case, the dimerization has been achieved by
alternating the

barrier width between 40 and 60~\AA; the well width is kept constant
at 90~\AA.

Such systems have interesting properties for  selectively enhanced
generation

of harmonic frequencies~\cite{ACS}. Fig.~7 shows a situation where
the two

bands do not cross, but avoid each other, so that a forbidden zone
appears

where, without level repulsion, the two bands would overlap.

This band--band repulsion gives rise to the appearance of an almost

degenerate pair of new edge states, with quasienergies in the
otherwise

forbidden zone. These localized states disappear when the amplitude
is

increased beyond the repulsion regime.

Since one of the most significant experimental features of
superlattices in
intense far--infrared laser fields~\cite{GUI} is their electrical
conductivity, a particularly important theoretical result

is the strong dependence of the localization length of defect states
on the
laser parameters.  As is well known~\cite{LEE}, {\it all} electronic
states are localized in a disordered one dimensional system, but the
conductivity is strongly affected when the localization length of the
relevant states becomes comparable to or less than the system size.
Though we have not looked explicitly here at more than a single
isolated defect, it is clear that a finite concentration of
disordered sites will also lead to states whose localization length
is sensitive to the amplitude and frequency of the applied laser
field.  This is a separate phenomenon from dynamical
localization~\cite{DLO} as a function of field amplitude or
frequency.  It has been shown~\cite{IGN} that the current in
ac--driven
ideal superlattices is proportional to $
J_{0}(e{\cal{E}}_{0}a/\omega) $, so
that it should vanish at the zeros of $ J_{0} $.  At these points the
quasi-energy spectrum is dispersionless; the group velocity is zero,
and
an initially localized wave packet does not spread,
but stays localized forever. The Floquet

states  are extended, but the individual components of a wave packet

can not dephase if all the their quasienergies are equal~\cite{QWS}.
In marked contrast, disorder leads to spatial localization of the
Floquet states

themselves, with a localization length that depends on the external
field parameters, and can be extremely short (and always {\it is} at
the same special points where  $ J_{0}(e{\cal{E}}_{0}a/\omega) = 0
$).

The conductivity thus depends on external field parameters for at
least two reasons. It will be interesting to study ac--driven
superlattices with
intentionally introduced disorder from this point of view.

This work was supported in part by  the

Alexander von Humboldt Foundation and by the Office of Naval Research
under Grant No. N00014-92-J-1452.

\vfill
\break

{\large{Figure captions}}

\vspace{5mm}

\begin{description}
\item[Fig.~1] Lowest quasienergy band for a model potential
consisting
	of 50 square wells with a width of 100~\AA, separated by
barriers

	50~\AA~wide and 0.3~eV high. The particle mass is
$0.067~m_{e}$,
	and the ac frequency is $ \omega = 10.0$~meV.
\item[Fig.~2] Lowest quasienergy band for a system with only 10
wells;
	the parameters are as in Fig.~1.
\item[Fig.~3] Lowest quasienergy band for a lattice with parameters
as in
	Fig.~1, but the width of the 24$^{th}$ barrier has been
decreased
	from 50 to 40~\AA.

	The ac frequency is $ \omega = 5.0$~meV.
\item[Fig.~4] Lowest quasienergy band for a lattice with parameters
as in
	Fig.~1, but the 24$^{th}$ well has been narrowed from 100 to
98~\AA.
	The ac frequency is $ \omega = 5.0$~meV.
\item[Fig.~5] Probability density of the Floquet state localized in
the

	neighborhood of a
	defect in a lattice with parameters as in Fig.~1; the ac
frequency
	is $ \omega = 5.0$~meV.
	The defect was induced by narrowing the 24$^{th}$ well by
0.5~\AA.

	The displayed interval in space covers a total of 10~wells;

	the interval in time is one laser period. The laser field
amplitude
	is

	(a) $ {\cal{E}}_{0} = 3000$~V/cm;
	(b) $ {\cal{E}}_{0} = 8016.08$~V/cm, the field strength of
the first
	quasienergy band collapse ($J_0 = 0$).
	The maximum density in each case is in the defect well.
\item[Fig.~6] Lowest doublet of quasienergy bands for a dimerized
system
	with a constant barrier width of 40~\AA~and well widths that
	alternate between 90 and 110~\AA. The barrier height and
particle
	mass are as in Fig.~1; the ac frequency is $ \omega =
2.0$~meV.
\item[Fig.~7] Anticrossing of two quasienergy bands in a dimerized
system
	where the well width is kept constant at 90~\AA, and the
barrier
	width alternates between 40 and 60~\AA. The ac frequency is
	$ \omega = 2.0$~meV.
\end{description}


\begin{thebibliography}{99}
\bibitem{fnote} The approximation is better than this estimate
suggests, since the integrand (and the integral) on the left hand
side oscillates about zero; for $t\rightarrow T$ the integral is
identically zero.
\end{thebibliography}
\end{document}